\documentclass[useAMS,usenatbib,fleqn]{mn2e}

\usepackage[draft]{hyperref}
\usepackage{amssymb}
\usepackage{amsmath}
\usepackage{mathtools}
\usepackage{enumitem}
\usepackage{graphicx}

\usepackage{nicefrac}
\usepackage{bigints}

\usepackage{color}

\newcommand{\firstit}[1]{\textcolor{black}{#1}}
\newcommand{\updated}[1]{\textcolor{black}{#1}}
\newcommand{\referee}[1]{\textcolor{black}{#1}}
\newcommand{\refereetwo}[1]{\textcolor{black}{#1}}

\newcommand{\rv}{\mbox{$R_\mathrm{V}$}}
\newcommand{\ebv}{\mbox{$E(B-V)$}}

\newcommand{\dmb}{\mbox{$\Delta M_{B,15}$}}
\newcommand{\dmbobs}{\mbox{$\Delta M_{B,15}^\mathrm{obs}$}}
\newcommand{\dmbtrue}{\mbox{$\Delta M_{B,15}^\mathrm{true}$}}
\newcommand{\ebvdlos}{\mbox{$E(B-V)_\mathrm{DLOS}$}}
\newcommand{\lmc}{\mbox{LMC$-$}}
\newcommand{\mw}{\mbox{MW$-$}}

\title[\referee{Shedding light on} the SN Ia extinction puzzle]{\referee{Shedding light on} the Type Ia supernova extinction puzzle: dust location found}
\author[M. Bulla et al.]{M.~Bulla,\thanks{E-mail: mattia.bulla@fysik.su.se} A. Goobar and S. Dhawan\\
Oskar Klein Centre, Department of Physics, Stockholm University, SE 106 91 Stockholm, Sweden
}

\date{Accepted 2018 June 14. Received 2018 June 5; in original form 2018 March 26.}

\begin{document}

\maketitle 

\begin{abstract}
The colour evolution of reddened Type Ia supernovae can place strong constraints on the location of dust and help address the question of whether the observed extinction stems from the interstellar medium or from circumstellar material surrounding the progenitor. Here we analyse BV photometry of 48 reddened Type Ia supernovae from the literature and estimate the dust location from their $B-V$ colour evolution. We find a time-variable colour excess \ebv~for 15 supernovae in our sample and constrain dust to distances between 0.013 and 45~pc ($4\times10^{16}-10^{20}$~cm). For the remaining supernovae, we obtain a constant \ebv~evolution and place lower limits on the dust distance from the explosion. In all the 48 supernovae, the inferred dust location is compatible with an interstellar origin for the extinction. \firstit{This is corroborated by the observation that supernovae with relatively nearby dust ($\lesssim$~1~pc) are located close to the center of their host galaxy}, in high-density dusty regions where interactions between the supernova radiation and interstellar clouds close by are likely to occur. For supernovae showing time-variable \ebv, we identify a potential preference for low \rv~values, unusually strong sodium absorption and blue-shifted and time-variable absorption features. Within the interstellar framework, this brings evidence to a proposed scenario where cloud-cloud collisions induced by the supernova radiation pressure can shift the grain size distribution to smaller values and enhance the abundance of sodium in the gaseous phase. 
\end{abstract}

\section{Introduction}
\label{sec:introduction}

In the past couple of decades, Type Ia supernovae (SNe~Ia) have been used as cosmological distance indicators and led to the discovery of the accelerated expansion of the Universe \citep{riess1998,perlmutter1999}. A key to such breakthrough discovery was the realization that the absolute brightness dispersion of SNe~Ia can be reduced by exploiting two main empirical relations. The first is the well-known ``width-luminosity'' relation, which corrects for the fact that intrinsically brighter SNe~Ia decline more slowly after peak \citep{phillips1993}. The second one, usually referred to as the ``colour-brightness'' relation, accounts for the fact that fainter SNe~Ia are typically redder than brighter SNe~Ia \citep{tripp1998}. Despite their importance for cosmological studies, the physical origin of these two relations has not yet been settled (see \citealt{goobar2011} for a review).

Intervening dust along the line of sight to the observer can dim the SN radiation and make it redder, thus possibly explaining the observed ``colour-brightness'' relation. However, both sample studies \citep{tripp1998,astier2006,nobili2008,kowalski2008,burns2014} and the analysis of individual SNe \citep{krisciunas2006,eliasrosa2006,eliasrosa2008,wang2008a,folatelli2010,amanullah2014,amanullah2015} clearly indicate that the extinction seen in many SNe~Ia is peculiar compared to the one in our Galaxy. Specifically, its wavelength dependence (the so-called extinction law) is much steeper than the one observed in the Milky Way (MW). This is usually parametrised by the total-to-selective extinction ratio \mbox{$R_\mathrm{V}=A_\mathrm{V}/(A_\mathrm{B}-A_\mathrm{V})\equiv A_\mathrm{V}/\ebv$}, where $A_\mathrm{X}$ is the extinction in the band $X$, with increasingly steeper extinction laws corresponding to increasingly smaller \rv. While $R_\mathrm{V}\sim3$ is typical for dust in our Galaxy \referee{\citep{cardelli1989}}, much lower values are \refereetwo{often} found in SNe~Ia \referee{($R_\mathrm{V}\lesssim2$, see references above)}. Independent confirmations of low \rv~come from spectropolarimetry of some reddened SNe~Ia \citep{patat2015,zelaya2017}.

The relatively low \rv~values have been taken as evidence for circumstellar (CS) material surrounding the progenitor at the time of explosion \citep{wang2005,patat2006,goobar2008}. In particular, multiple-scattering of photons within a CS shell can increase the relative extinction at shorter wavelengths and thus steepen the resulting extinction law. The CS scenario would be consistent with the claimed detection of CS gas from blue-shifted and time-variable sodium and potassium features \citep{patat2007,simon2009,sternberg2011,
foley2012,maguire2013,graham2015,ferretti2016} and with the observed similarities between the continuum polarization of some reddened SNe~Ia and that of proto-planetary nebulae \citep{cikota2017}. 

\begin{table}
\centering
\caption{Sample of reddened SNe~Ia used in this study. Properties extracted with \textsc{snoopy} \citep{burns2014} and references for photometry are reported for each SN.}
\label{tab:sne}
\begin{tabular}{lcccc}
\hline
SN & $t_\mathrm{max}(B)$ & \dmbobs & $E(B-V)_\mathrm{MW}$ & Ref. \\
 & (MJD) & (mag) & (mag) & \\
\hline
SN~1986G  & 46560.79 & 1.66 & 0.108 & 1 \\[0.027cm]
SN~1989B  & 47563.84 & 1.09 & 0.028 & 2 \\[0.027cm]
SN~1995E	 & 49774.00 & 0.98 & 0.023 & 3 \\[0.027cm]
SN~1996ai & 50255.30 & 0.99 & 0.012 & 3 \\[0.027cm]
SN~1996bo & 50386.24 & 1.21 & 0.067 & 3 \\[0.027cm]
SN~1997dt & 50785.83 & 1.15 & 0.048 & 4 \\[0.027cm]
SN~1998bu & 50952.30 & 1.02 & 0.022 & 5 \\[0.027cm]
SN~1998dm & 51060.14 & 0.86 & 0.038 & 6 \\[0.027cm]
SN~1999cl & 51341.80 & 1.09 & 0.032 & 6 \\[0.027cm]
SN~1999ee & 51468.32 & 0.82 & 0.017 & 7 \\[0.027cm]
SN~1999gd & 51519.61 & 1.27 & 0.035 & 4 \\[0.027cm]
SN~2000ce & 51665.24 & 0.93 & 0.047 & 4 \\[0.027cm]
SN~2000cp & 51722.21 & 1.26 & 0.042 & 6 \\[0.027cm]
SN~2001E	 & 51925.88 & 0.97 & 0.033 & 6 \\[0.027cm]
SN~2001dl & 52130.90 & 0.91 & 0.046 & 6 \\[0.027cm]
SN~2002G & 52299.02 & 1.38 & 0.011 & 6 \\[0.027cm]
SN~2002bo & 52356.79 & 1.14 & 0.022 & 6 \\[0.027cm]
SN~2002cd & 52383.32 & 0.91 & 0.350 & 6 \\[0.027cm]
SN~2002hw & 52595.81 & 1.39 & 0.093 & 8 \\[0.027cm]
SN~2002jg & 52609.82 & 1.39 & 0.057 & 6 \\[0.027cm]
SN~2003cg & 52729.62 & 1.09 & 0.026 & 8 \\[0.027cm]
SN~2003hx & 52892.50 & 1.44 & 0.072 & 9 \\[0.027cm]
SN~2004ab & 53056.32 & 1.17 & 0.070 & 10\\[0.027cm]
SN~2005A	 & 53379.58 & 1.14 & 0.026 & 11 \\[0.027cm]
SN~2005bc & 53470.33 & 1.37 & 0.009 & 6 \\[0.027cm]
SN~2005kc & 53698.04 & 1.25 & 0.114 & 8 \\[0.027cm]
SN~2006X	 & 53785.62 & 1.11 & 0.023 & 11 \\[0.027cm]
SN~2006br & 53853.89 & 1.53 & 0.020 & 8,11 \\[0.027cm]
SN~2006cc & 53873.83 & 0.91 & 0.011 & 8 \\[0.027cm]
SN~2006cm & 53885.48 & 0.91 & 0.041 & 8 \\[0.027cm]
SN~2006gj & 53999.94 & 1.55 & 0.070 & 11 \\[0.027cm]
SN~2006os & 54067.08 & 1.27 & 0.125 & 11 \\[0.027cm]
SN~2007S	 & 54143.29 & 0.80 & 0.022 & 11 \\[0.027cm]
SN~2007bm & 54224.58 & 1.17 & 0.035 & 8 \\[0.027cm]
SN~2007ca & 54226.68 & 0.77 & 0.057 & 11 \\[0.027cm]
SN~2007cg & 54230.62 & 1.06 & 0.069 & 8,11 \\[0.027cm]
SN~2007cs & 54275.66 & 0.80 & 0.057 & 12 \\[0.027cm]
SN~2007le & 54398.74 & 0.95 & 0.029 & 11 \\[0.027cm]
SN~2007ss & 54452.55 & 1.12 & 0.013 & 12 \\[0.027cm]
SN~2008dt & 54646.04 & 0.93 & 0.041 & 6,12 \\[0.027cm]
SN~2008fp & 54730.52 & 0.92 & 0.169 & 11 \\[0.027cm]
SN~2009I & 54851.94 & 0.87 & 0.027 & 11 \\[0.027cm]
SN~2009fv & 55000.44 & 1.76 & 0.006 & 12 \\[0.027cm]
SN~2010ev & 55384.51 & 1.14 & 0.089 & 13 \\[0.027cm]
SN~2012bm & 56018.00 & 0.72 & 0.010 & 14 \\[0.027cm]
SN~2012cp & 56081.10 & 0.86 & 0.018 & 14 \\[0.027cm]
SN~2012cu & 56105.16 & 0.90 & 0.022 & 14 \\[0.027cm]
SN~2014J  & 56689.64 & 0.96 & 0.050 & 15 \\[0.027cm]
\hline
\end{tabular}
\begin{flushleft}
\textbf{Photometry References}. (1) \citealt{phillips1987}; (2) \citealt{wells1994}; (3) \citealt{riess1999}; (4) \citealt{jha2006}; (5) \citealt{jha1999}; (6) \citealt{ganeshalingam2010}; (7) \citealt{stritzinger2002}; (8) \citealt{hicken2009}; (9) \citealt{misra2008}; (10) \citealt{chakradhari2018}; (11) \citealt{krisciunas2017}; (12) \citealt{hicken2012}; (13) \citealt{gutierrez2016}; (14) \citealt{amanullah2015}; (15) \citealt{srivastav2016}. 
\end{flushleft}
\end{table}

Another possible interpretation is that the low \rv~values are caused by dust in the interstellar (IS) medium. This would require IS clouds in the SN host galaxies to be characterized by smaller dust grains compared to those seen in our Galaxy \citep{gao2015,nozawa2016}, possibly as a consequence of cloud-cloud collisions induced by the SN radiation pressure \citep{hoang2017}. Evidences supporting the IS scenario were given by \citet{phillips2013} who found a strong correlation between the extinction of SNe~Ia and the strength of a diffuse interstellar band at 5870~\AA. In addition, a recent study by \citet{bulla2018} demonstrated how the colour evolution of reddened SNe~Ia can put strong constraints on the location of dust. In particular, the time-evolution of the colour excess \ebv~in SN~2006X and SN~2014J was found to be consistent with dust in the IS medium.

Understanding the origin of dust extinction is also important to constraint the progenitor system(s) of SNe~Ia since the amount of CS material is predicted to vary among different scenarios. While SNe~Ia are widely-accepted to arise from thermonuclear explosion of carbon-oxygen white dwarves in binary systems, whether the companion star is a non-degenerate star (``single-degenerate'', SD, scenario) or another white dwarf (``double-degenerate'', DD, scenario) is still hotly debated (see e.g. \citealt{livio2018} and references therein). In the SD scenario, the high mass transfer preceding the explosion creates high-density CS material with which the SN radiation is likely to interact. In contrast, a cleaner local environment is expected from DD models. 

In this paper, we use the technique introduced by \citet{bulla2018} and estimate dust locations in 48 SNe~Ia from the literature. We describe our dataset in Section~\ref{sec:data} and outline the approach used to estimate \ebv~for both models and observations in Section~\ref{sec:extinction}. We then present our results in Section~\ref{sec:results} and discuss their implication for the CS/IS debate in Section~\ref{sec:discussion}. Finally, we draw conclusions in Section~\ref{sec:conclusions}.

\section{Data}
\label{sec:data}

Photometric data of the 48 SNe~Ia analysed in this work are taken from the literature. In particular, photometry for 38 SNe is obtained from either the CfA supernova program \citep{riess1999,jha1999,jha2006,hicken2009,hicken2012}, the Lick Observatory Supernova Search (LOSS, \citealt{ganeshalingam2010}) or the Carnegie Spectroscopic Project (CSP, \citealt{krisciunas2017}). Photometry references for all the SNe are reported in Table~\ref{tab:sne}. The time of maximum in the $B-$band, $t_\mathrm{max}(B)$, the decline in the $B-$band magnitude from peak to 15~d after, \dmbobs, and the MW colour excess, $E(B-V)_\mathrm{MW}$, are calculated using \textsc{snoopy} \citep{burns2014} and \referee{also} reported for each SN in Table~\ref{tab:sne}.

\begin{figure}
\begin{center}
\includegraphics[width=1\columnwidth]{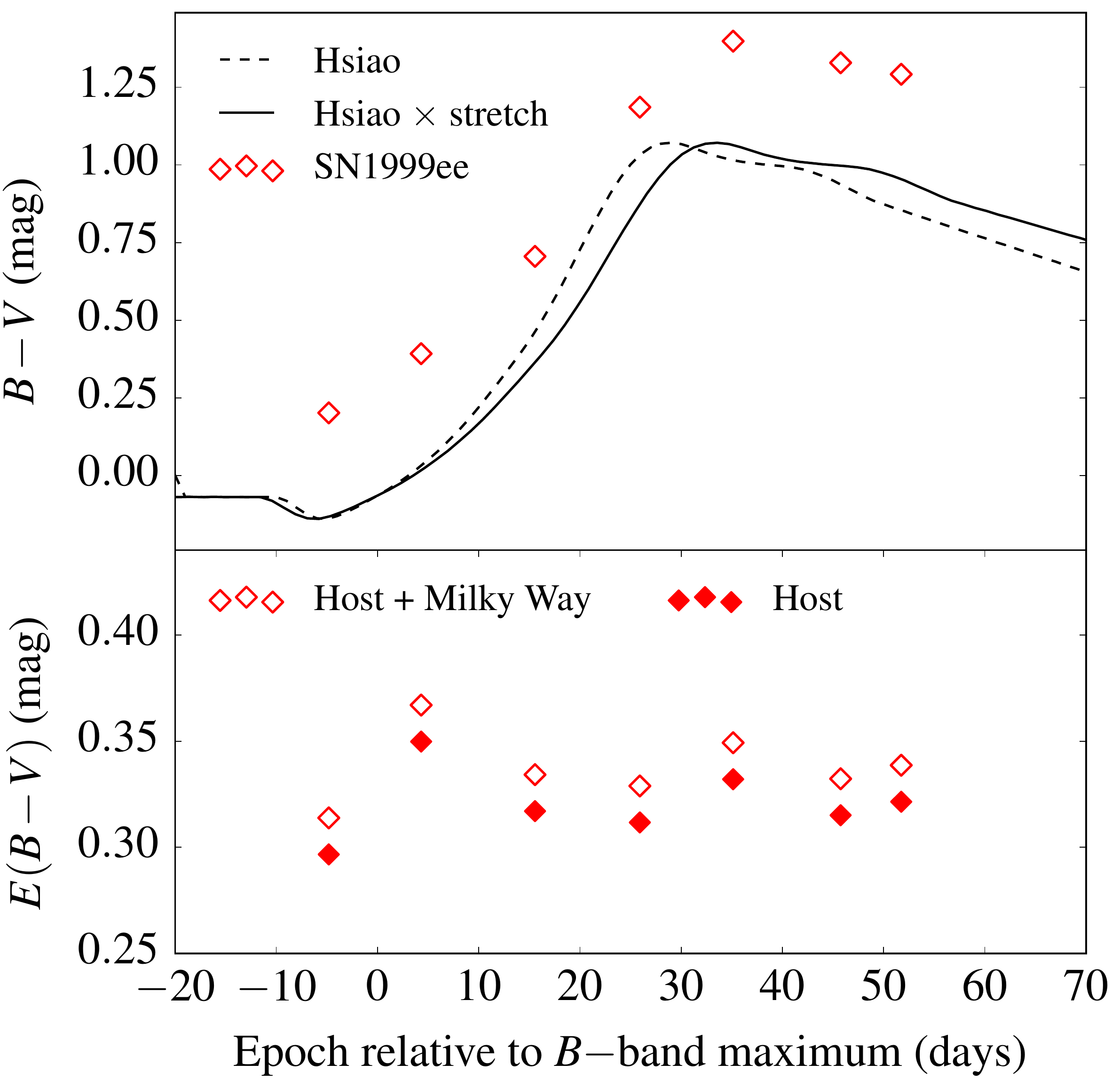}
\caption{Procedure to extract \ebv~values from observations. The specific case of SN~1999ee is shown. In the upper panel, $B-V$ colours as a function of time are shown for SN~1999ee (diamonds) and for the Hsiao template before (dashed line) and after (solid line) the stretch in time has been applied. In the bottom panel, \ebv~values as a function of time are shown before (open diamonds) and after (filled diamonds) subtracting the MW colour excess, \ebv$_\mathrm{MW}$. }
\label{fig:procedure}
\end{center}
\end{figure}

\section{Estimating colour excess}
\label{sec:extinction}

In this Section, we outline the approach use throughout the paper to estimate colour excess values \ebv.
As discussed below, extracting colour excess values from both dust models (Section~\ref{sec:models}) and observations (Section~\ref{sec:obs}) require some assumptions on the intrinsic SN light curves. For most of the SNe in our sample, the pristine light curves are constructed from \referee{the spectral template of} \citet{hsiao2007}. For SN~2003hx, SN~2007cf and SN~2014J, photometric data \referee{that are relevant to capture the full \ebv~evolution and to provide strong constraints on the dust localization} extend to epochs not covered by the Hsiao template ($\gtrsim85$~d)\referee{. For these particular cases,} we take the low-extinction \citep{patat2013}, \referee{apparently normal} `plain vanilla' \citep{wheeler2012} SN~2011fe as representative of a pristine SN~Ia. In particular, we adopt photometric data of SN~2011fe from \citet{munari2013} as these extend to 325~d after peak. 

\refereetwo{Here we stress that SN~2011fe is used only when necessary to properly characterize the \ebv~evolution of the reddened SNe, while preferring a template constructed from a sample of SNe~Ia (as the Hsiao template is) for the analysis of most of the objects. We note, however, that the choice of the template is important, with templates extending to later epochs typically leading to stronger constraints on the dust localization (see discussion in Section~\ref{sec:results}).}

\subsection{Models}
\label{sec:models}

In this work, we use the Monte Carlo code described by \citet{bulla2018} and publicly available at \url{https://github.com/mbulla/dust3d}. \firstit{The code is similar to those in \citet{goobar2008} and \citet{amanullah2011}}\referee{, and} simulates the propagation of Monte Carlo photons through a dust region. In particular, it keeps track of the accumulated time-delay of each photon compared to non-interacting photons, $\Delta t$. The dust region is modelled as a spherical shell at a given distance $\zeta$ from the SN, with an optical depth $\tau_\lambda$ set to give a desired colour excess along the direct line-of-sight (DLOS), \ebvdlos. \referee{For the range of \ebv~values used in this work, $0.3~\mathrm{mag}\le$~\ebvdlos~$\le1.75$~mag, optical depths in the $B-$ and $V-$band are in the range $1.2\le\tau_B\le6.8$ and $0.9\le\tau_V\le5.2$, respectively.} Dust properties are taken from \citet{draine2003} and corresponds to \mw type dust\footnote{In terms of \ebv~time evolution, our models are not very sensitive to the specific dust composition (see figs. 2 and 3 of \citealt{bulla2018} and discussion in Section~\ref{sec:lightechoes}).}. 

Photons escaping the dust shell are collected and used to construct a light curve after interaction with dust. As described by \citet{bulla2018}, this is done by using a SN template (see above) and taking into account the different time-delays $\Delta t$ of photons escaping the dust shell. When comparing models to data in Section~\ref{sec:results}, templates are stretched in time by a convenient amount $s$ so that \dmb~for the constructed light-curve matches the one observed for the SN under investigation, \dmbobs~(see Table~\ref{tab:sne}). At each time, the extinction $A_X$ in any given $X-$band is then computed as the difference between the template and the constructed light-curve, and the desired colour excess calculated as $\ebv=A_B-A_V$.

\subsection{Observations}
\label{sec:obs}

\begin{figure*}
\begin{center}
\includegraphics[width=1\textwidth]{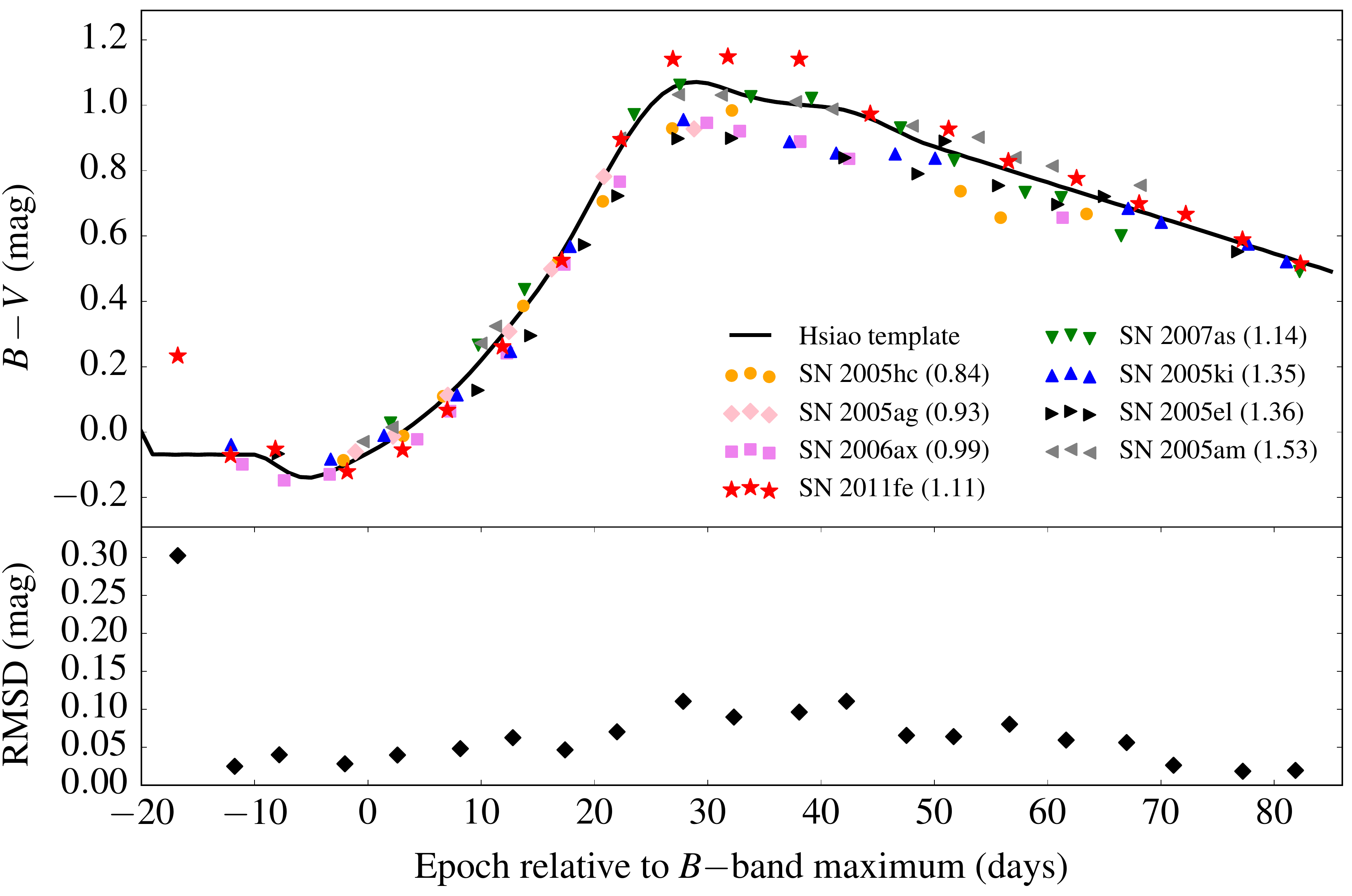}
\caption{Estimate of the intrinsic $B-V$ colour variations. In the upper panel, $B-V$ colours are shown for 8 low-reddened SNe~Ia (points) after stretching their light curves to match \dmb~of the Hsiao template (black line) \firstit{and subtracting the extragalactic and Galactic \ebv~values reported in \citet{burns2014}}. In the bottom panel, the root-mean-square deviations (RMSD) of the 8 SNe about the Hsiao template are shown in 5-day windows from $-$20 to +85~d relative to $B-$band maximum. RMSD values thus calculated are taken as representative of the intrinsic SN colour variations and are reported in Table~\ref{tab:intrinsic}. }
\label{fig:intrinsic}
\end{center}
\end{figure*}

For the SNe in our sample, \ebv~are calculated by comparing the observed $B-V$ colours to those from the adopted template. The procedure is shown in Fig.~\ref{fig:procedure} for the specific case of SN~1999ee and summarized in the following steps:
\begin{itemize}[leftmargin=*]
\item Photometric data points are averaged over 10-day windows following \citet{bulla2018}; 
\item The SN template is stretched in time by a factor $s$ (see above);
\item $B-V$ colours are calculated for both the observations and the template;
\item The total (host + MW) colour excess is calculated by subtracting the template $B-V$ from the observed $B-V$; \item The MW colour excess of the SN (Table~\ref{tab:sne}) is subtracted to find the desired \firstit{extragalactic} colour excess \ebv.
\end{itemize}
This procedure is analogous to the one adopted in \citet{amanullah2014} and \citet{bulla2018} but does not require to fit photometric data for both \ebv~and $R_\mathrm{V}$. This is justified by the relatively small impact of different dust compositions (and thus corresponding $R_\mathrm{V}$) on the predicted colour excess values (see fig. 3 of \citealt{bulla2018} and discussion in Section~\ref{sec:lightechoes}). The different approaches were tested on SN~2006X and found to give similar results.

\begin{table}
\centering
\caption{Intrinsic SN variations in $B-V$ colour in different time ranges. Epochs are relative to $t_\mathrm{max}(B)$.}
\label{tab:intrinsic}
\begin{tabular}{rrc}
\hline
From & To & $\sigma_\mathrm{intr}$ \\
(d) & (d) & (mag)\\
\hline
$-20$ & $-15$ & 0.302\\
$-15$ & $-10$ & \updated{0.025}\\
$-10$ & $-5$ & \updated{0.040}\\
$-5$ & $0$ & \updated{0.028}\\
$0$ & $+5$ & \updated{0.039}\\
$+5$ & $+10$ & \updated{0.048}\\
$+10$ & $+15$ & \updated{0.063}\\
$+15$ & $+20$ & \updated{0.046}\\
$+20$ & $+25$ & \updated{0.070}\\
$+25$ & $+30$ & \updated{0.110}\\
$+30$ & $+35$ & \updated{0.090}\\
\hline
\end{tabular}
\begin{tabular}{rrc}
\hline
From & To & $\sigma_\mathrm{intr}$ \\
(d) & (d) & (mag)\\
\hline
$+35$ & $+40$ & \updated{0.096}\\
$+40$ & $+45$ & \updated{0.111}\\
$+45$ & $+50$ & \updated{0.065}\\
$+50$ & $+55$ & \updated{0.064}\\
$+55$ & $+60$ & \updated{0.080}\\
$+60$ & $+65$ & \updated{0.059}\\
$+65$ & $+70$ & \updated{0.056}\\
$+70$ & $+75$ & \updated{0.026} \\
$+75$ & $+80$ & \updated{0.018} \\
$+80$ & $+85$ & \updated{0.019} \\
 & & \\
\hline
\end{tabular}
\end{table}  

\begin{table*}
\centering
\caption{Dust distance values, $\zeta$, for the 15 SNe in our sample showing a time-variable \ebv. For each SN, the $\chi^2$ for the best-fit, $\chi^2_\nu$, is reported together with that from a model with a constant colour excess ($\zeta=+\infty$), $\chi^2_{\nu,\mathrm{const}}$. Inferred \ebvdlos~and \dmbtrue~are also given. SNe are ordered by increasing dust distance $\zeta$.}
\label{tab:results1}
\begin{tabular}{lcccccc}
\hline
SN & \ebvdlos~(mag) & $\zeta$ (pc) & $\chi^2_{\nu}$ & $\chi^2_{\nu,\mathrm{const}}$ & \dmbobs~(mag) & \dmbtrue~(mag) \\
\hline
SN~2003hx & 0.50 & 0.013$_{-0.001}^{+0.001}$ & 4.14 & \updated{22.68} & 1.36 & \updated{1.69$_{-0.04}^{+0.06}$} \\[0.17cm]
SN~1996bo & \updated{0.39} & 0.70\updated{$_{-0.04}^{+0.22}$} & \updated{4.12} & \updated{21.70} & 1.21 & 1.46\updated{$_{-0.06}^{+0.01}$} \\[0.17cm]
SN~2007cs & 1.38 & 1.0$_{-0.3}^{+0.3}$ & \updated{0.54} & \updated{19.66} & 0.80 & 1.66\firstit{$_{-0.25}^{+0.48}$} \\[0.17cm]
SN~2008dt & 0.62 & 1.0$_{-0.2}^{+0.6}$ & 0.26 & \updated{5.11} & 0.93 & 1.17\firstit{$_{-0.09}^{+0.05}$} \\[0.17cm]
SN~2006br & 1.00 & 1.7$_{-0.2}^{+0.4}$ & \updated{1.85} & \updated{14.81} & 1.53 & 1.85\firstit{$_{-0.06}^{+0.04}$} \\[0.17cm]
SN~2005A	& 1.04 & \updated{4.7$_{-0.8}^{+1.3}$} & \updated{2.31} & \updated{8.34} & 1.14 & \updated{1.27$_{-0.03}^{+0.03}$} \\[0.17cm] 
SN~2007S	 & \updated{0.42} & \updated{5.7$_{-1.0}^{+1.5}$} & \updated{0.22} & \updated{4.94} & 0.80 & 0.82\firstit{$_{-0.01}^{+0.01}$} \\[0.17cm]
SN~2007le & 0.34 & \updated{5.9$_{-1.0}^{+1.2}$} & \updated{0.29} & \updated{6.51} & 0.95 & 0.97\firstit{$_{-0.01}^{+0.01}$} \\[0.17cm]
SN~2008fp & 0.53 & \updated{9.4$_{-0.7}^{+1.1}$} & \updated{2.18} & \updated{22.25} & 0.92 & 0.94\firstit{$_{-0.01}^{+0.01}$} \\[0.17cm]
SN~2002cd & 0.70 & \updated{9.6$_{-1.7}^{+1.2}$} & \updated{1.36} & \updated{6.18} & 0.91 & 0.94\firstit{$_{-0.01}^{+0.01}$} \\[0.17cm]
SN~1999cl & \updated{1.08} & 12.8\updated{$_{-2.9}^{+4.9}$} & 0.49 & \updated{3.74} & 1.09 & 1.13\firstit{$_{-0.01}^{+0.01}$} \\[0.17cm]
SN~1986G  & \updated{1.00} & \updated{14.5$_{-1.4}^{+3.3}$} & \updated{1.27} & \updated{10.36} & 1.66 & 1.70\firstit{$_{-0.01}^{+0.01}$} \\[0.17cm]
SN~2002bo & 0.46 & \updated{15.9$_{-2.8}^{+3.5}$} & \updated{1.16} & \updated{4.26} & 1.14 & 1.15\firstit{$_{-0.01}^{+0.01}$} \\[0.17cm]
SN~2006X	& \updated{1.24} & 17.3\updated{$_{-1.0}^{+1.3}$} & \updated{1.14} & \updated{39.02} & 1.11 & 1.15\firstit{$_{-0.01}^{+0.01}$} \\[0.17cm]
SN~2014J  & 1.40 & 44.6\updated{$_{-2.0}^{+3.3}$} & \updated{1.11} & \updated{19.98} & 0.96 & 0.97\firstit{$_{-0.01}^{+0.01}$} \\
\hline                           
\end{tabular}                    
\end{table*}

Uncertainties on the extracted \ebv~includes propagation of errors on photometric data, $\sigma_\mathrm{data}$, and intrinsic SN colour variations, $\sigma_\mathrm{intr}$. The latter is estimated by comparing $B-V$ colours of individual low-reddening SNe~Ia from the CSP sample \citep{burns2014} to those predicted by the Hsiao template. \firstit{In particular, we follow \citet{folatelli2010} and use SNe that are either in E/S0-type galaxies, located away from dusty regions or with no detection of Na~I~D lines in early-time spectra}. This is shown in Fig.~\ref{fig:intrinsic} for 8 SNe~Ia with $\ebv<0.05$~mag \citep{burns2014} and with a range of observed \dmbobs~spanning the one in our sample of 48 reddened SNe~Ia ($0.8-1.6$~mag). After stretching light curves for the 8 SNe~Ia to match the same \dmb~of the Hsiao template, the root-mean-square deviations (RMSD) about the template are typically smaller than $\sim$~0.1~mag. \firstit{For epochs earlier than $-$15~d, instead, the RMSD is large (0.3 mag) and due to the difference in colour between SN~2011fe and the Hsiao template. Nonetheless, we note that this high RMSD has no impact on our analysis since no SN in our sample has available photometry at these early epochs}. The inferred RMSD values are taken as representative of the intrinsic SN colour variations and are reported in Table~\ref{tab:intrinsic}. These are in good agreement with those estimated by \citet{nobili2008}. \firstit{We notice that the intrinsic $B-V$ colour variations reach $\sim$~0.1~mag at $\sim30-40$~d after $B-$band maximum (when the $B-V$ curve is at its peak)} but they are typically smaller ($\lesssim$~0.05~mag) at earlier and later epochs. This point will be important in Section~\ref{sec:results} when constraining the dust distance from the \ebv~temporal evolution of some SNe~Ia.

\section{Results}
\label{sec:results}

\firstit{As described by \citet{bulla2018}, no temporal variations in the colour excess are expected for dust relatively distant from the SN since the percentage of photons removed from the DLOS by dust -- and thus the extinction -- is the same at all epochs. In contrast, a time-variable colour excess is expected for relatively nearby dust due to the simultaneous arrival to the observer of non-interacting photons and of photons scattered by dust. Compared to non-interacting photons, scattered photons have travelled longer and were thus emitted earlier with a different brightness and colour. As a result of the intrinsic SN colour evolution (see Fig.~\ref{fig:intrinsic}), scattered photons are typically bluer than non-interacting photons at early epochs, while redder than non-interacting photons at late epochs. Therefore -- after an initial phase of constant colour excess when the contribution of scattered photons is negligible -- the colour excess evolution is characterized by a first drop (i.e. decreasing reddening) and then by a later rise (i.e. increasing reddening)}.

In this section, we analyse the 48 SNe~Ia in our sample and infer dust distance values through the comparison between observed and predicted \ebv. We find 15 SNe to be characterized by a time-variable colour excess~(Section~\ref{sec:estimates}), while the remaining 33 SNe to be consistent with a constant colour-excess evolution (Section~\ref{sec:lower}). Following \citet{bulla2018}, we extract distance values for the first sub-sample while distance lower limits for the second sub-sample.

Best-fit dust models, and corresponding most-probable distances $\zeta$, are identified via a simple $\chi^2$ test. The extracted dust distance values, together with the inferred \ebvdlos~and \dmbtrue, are summarized in Table~\ref{tab:results1} and Table~\ref{tab:results2}. For SNe showing a time-variable \ebv, we report the most-probable distance values and their 1$\sigma$-uncertainties; for SNe compatible with a constant \ebv~evolution, instead, we report 3$\sigma$ lower-limits on the dust distance.

\subsection{Dust distance estimates}
\label{sec:estimates}

\begin{figure*}
\begin{center}
\includegraphics[width=1\textwidth]{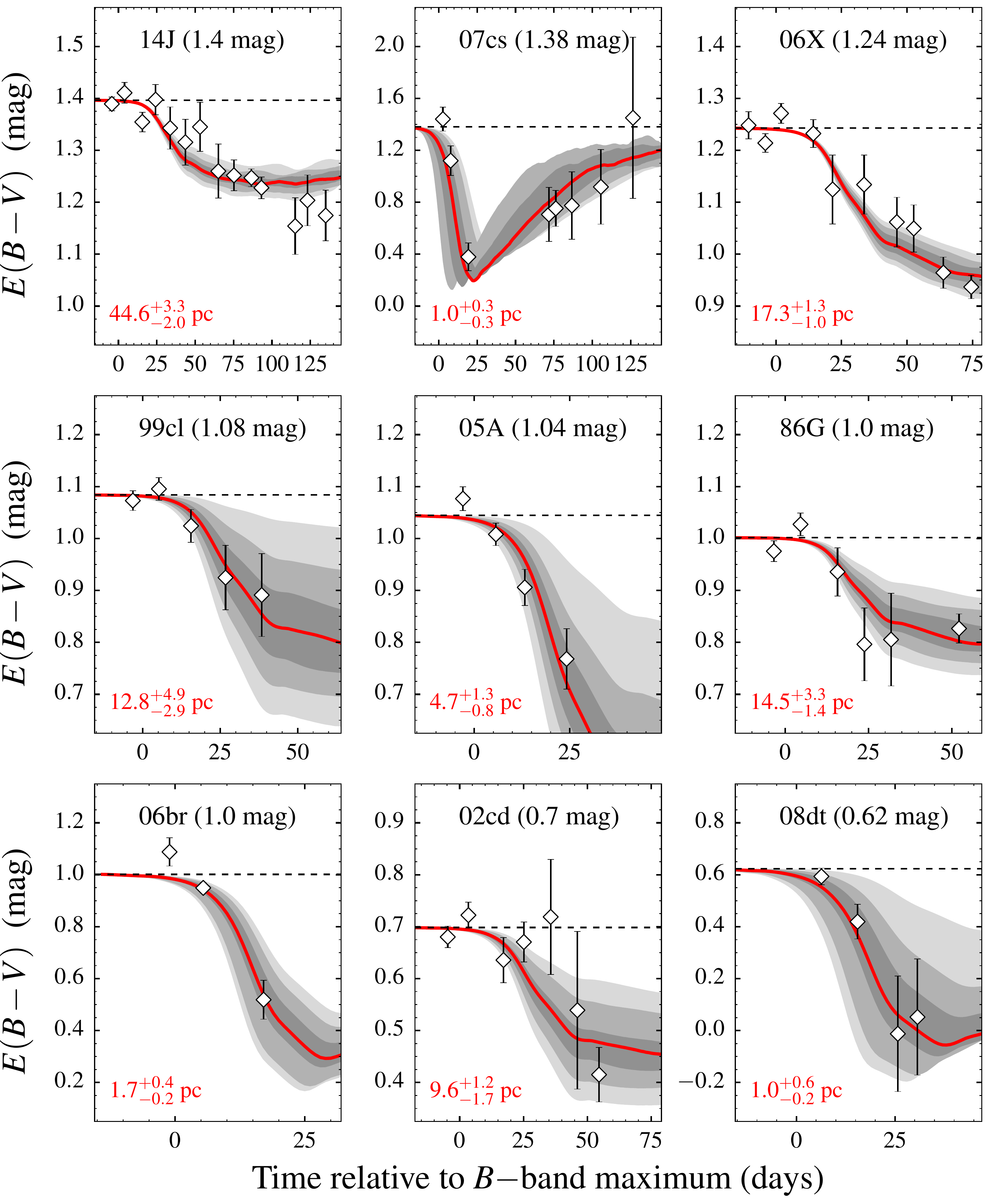}
\caption{\ebv~evolution for 9 of the 15 SNe~Ia showing time-variable extinction (white diamonds). SNe are ordered with decreasing \ebvdlos~(from $1.4$ to $0.62$~mag) from top-left to bottom-right panels. Solid red lines represent best-fit dust models for each SN, while grey areas the 68.3 $\%$ (1$\sigma$), 95.4 $\%$ (2$\sigma$) and 99.7 $\%$ (3$\sigma$) statistical confidence contours (from darker to lighter). The inferred distances are reported in the bottom-left corner of each panel, together with their 1$\sigma$ uncertainties. \firstit{Horizontal dashed lines in each panel mark the constant \ebv~evolution for $\zeta=+\infty$}. Note that $x-$axes have different scales in different panels, while $y-$axes are the same in all panels (0.65~mag) except for SN~2007cs (2.6~mag, four times as large), SN~2006br and SN~2008dt (1.3~mag, twice as large).}
\label{fig:results1}
\end{center}
\end{figure*}

\begin{figure*}
\begin{center}
\includegraphics[width=1\textwidth]{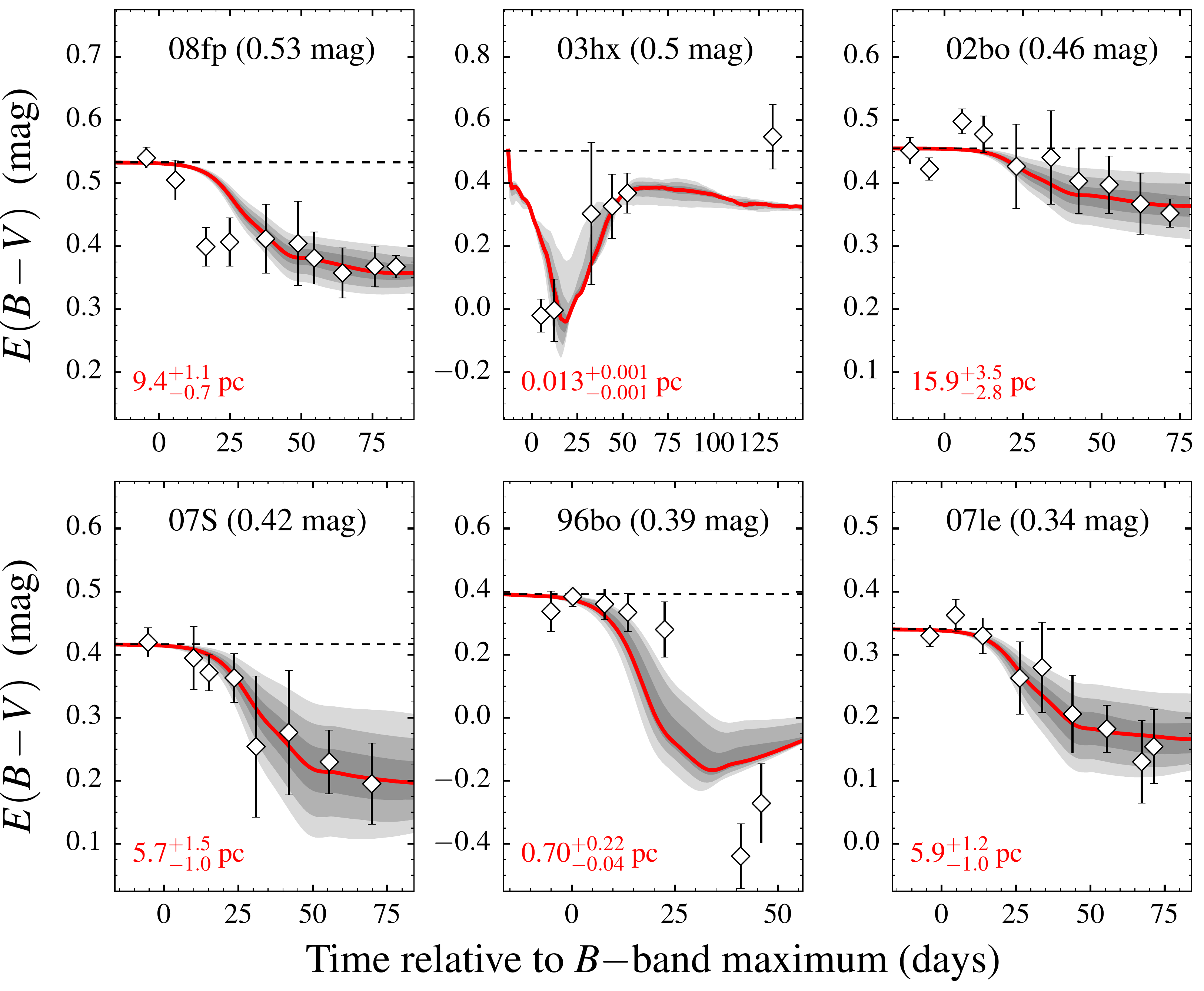}
\caption{\ebv~evolution for 6 of the 15 SNe~Ia showing time-variable extinction (white diamonds). SNe are ordered with decreasing \ebvdlos~(from $0.53$ to $0.34$~mag) from top-left to bottom-right panels. Solid red lines represent best-fit dust models for each SN, while grey areas the 68.3 $\%$ (1$\sigma$), 95.4 $\%$ (2$\sigma$) and 99.7 $\%$ (3$\sigma$) statistical confidence contours (from darker to lighter). The inferred distances are reported in the bottom-left corner of each panel, together with their 1$\sigma$ uncertainties. \firstit{Horizontal dashed lines in each panel mark the constant \ebv~evolution for $\zeta=+\infty$}. Note that $x-$axes have different scales in different panels, while $y-$axes are the same in all panels (0.65~mag) except for SN~2003hx and SN1996bo (1.3~mag, twice as large).}
\label{fig:results2}
\end{center}
\end{figure*}

Fig.~\ref{fig:results1} and Fig.~\ref{fig:results2} show the extracted \ebv~values for the 15~SNe~Ia with time-variable extinction, together with the inferred best-fit dust models.
The 15 SNe span a wide range of \ebvdlos, going from the highly-reddened SN~2014J (1.40~mag) to SN~2007le (0.34~mag).

For 13 of the 15 SNe, we are able to capture only the initial constant phase and the drop in \ebv. This is because \refereetwo{the rise in \ebv~predicted in our models is not captured either by the available photometry or by the adopted template (see discussion in Section~\ref{sec:extinction})}. The inferred locations of dust for these SNe span a range between 0.7 (SN~1996bo) and $\sim$~45~pc (SN~2014J). The full evolution is instead captured by the remaining two SNe. In the case of SN~2007cs, this is due to the combination of relatively nearby dust ($\sim$~1~pc) and good time-sampling of the light curve (extending to $\sim150$~d~after peak). In the case of SN~2003hx, instead, the dust is quite nearby the explosion site and thus the \ebv~evolution is faster. Specifically, our models suggest that dust is located at 0.013~pc. In Section~\ref{sec:intertellar}, we discuss the possibility that the relatively nearby dust toward this SN is suggestive of CS material connected to the progenitor system.

We note that the distance locations estimated here for SN~2006X and SN~2014J differ from those in \citet{bulla2018}. This is due to the use of different photometric data for these SNe. For SN~2006X, we adopt data from \citet{krisciunas2017} and obtain a distance value of \updated{$\zeta=17.3_{-1.0}^{+1.3}$}~pc, \referee{slightly smaller than} the value obtained by \citet{bulla2018} using data from \citet{wang2008a}, $\zeta=22.1_{-3.8}^{+6.0}$~pc. For SN~2014J, we use photometry from \citet{srivastav2016} extending to $\sim150$~d after maximum, that is to much later epochs than those covered by photometry from \citet[][$\lesssim40$~d~post-peak]{amanullah2015} used in \citet{bulla2018}. While a roughly constant \ebv~was found in \citet{bulla2018} and only a $1\sigma$ lower limit\footnote{Note that $1\sigma$ lower limits were reported in \citealt{bulla2018}, while here we adopt a more appropriate and conservative choice of $3\sigma$ for the lower limits (see Table~\ref{tab:results2}).} of $\zeta>38$~pc was placed on the dust distance, here we identify a drop in \ebv~starting from $\sim$~40~d and constraint the presence of dust at \updated{$\zeta=44.6_{-2.0}^{+3.3}$}. This demonstrates the importance of late-time photometry to constraint dust distances with our technique.

\begin{table*}
\centering
\caption{Dust distance lower limits, $\zeta$, for the 33 SNe in our sample showing a constant \ebv. For each SN, the $\chi^2$ for the $3\sigma$ lower-limit model, $\chi^2_\nu$, is reported together with that from a model with a constant colour excess ($\zeta=+\infty$), $\chi^2_{\nu,\mathrm{const}}$. Inferred \ebvdlos~and \dmbtrue~are also given. SNe are ordered by dust distance lower limits.}
\label{tab:results2}
\begin{tabular}{lcccccc}
\hline
SN & \ebvdlos~(mag) & $\zeta$ (pc) & $\chi^2_\nu$ & $\chi^2_{\nu,\mathrm{const}}$ & \dmbobs~(mag) & \dmbtrue~(mag) \\
\hline
SN~2000cp & 0.51 & $>$ 0.3 & \updated{7.83} & \updated{1.66} & 1.26 & 1.26$-$1.57 \\[0.11cm]
SN~2006cc & \updated{0.44} & $>$ \updated{0.3} & \updated{6.78} & \updated{5.74} & 0.91 & 0.91$-$\updated{1.13} \\[0.11cm]
SN~2009I & \updated{0.70} & $>$ 0.4 & \updated{5.85} & \updated{5.63}  & 0.87 & 0.87$-$1.21 \\[0.11cm]
SN~2012bm & 0.38 & $>$ 0.4 & \updated{8.17} & 0.01 & 0.72 & 0.72$-$0.84 \\[0.11cm]
SN~2009fv & 0.61 & $>$ \updated{0.5} & \updated{5.80} & \updated{1.17} & 1.76 & 1.76$-$\updated{1.98} \\[0.11cm]
SN~2007ca & \updated{0.29} & $>$ 0.7 & \updated{3.02} & \updated{1.79} & 0.77 & 0.77$-$0.82 \\[0.11cm]
SN~2007cg & 0.76 & $>$ 0.7 & \updated{5.06} & \updated{0.92} & 1.06 & 1.06$-$1.32 \\[0.11cm]
SN~2002jg & 0.69 & $>$ \updated{0.8} & \updated{4.76} & \updated{1.93} & 1.39 & 1.39$-$\updated{1.58} \\[0.11cm]
SN~2002hw & \updated{0.55} & $>$ \updated{0.9} & \updated{3.58} & \updated{1.80} & 1.39 & 1.39$-$\updated{1.55} \\[0.11cm]
SN~2001E	 & 0.52 & $>$ 0.9 & \updated{2.87} & \updated{0.78} & 0.97 & 0.97$-$1.07 \\[0.11cm]
SN~2006os & 0.34 & $>$ 1.0 & \updated{5.26} & \updated{2.44} & 1.27 & 1.27$-$1.33 \\[0.11cm]
SN~2002G & 0.39 & $>$ 1.1 & \updated{3.50} & \updated{0.69} & 1.38 & 1.38$-$1.47 \\[0.11cm]
SN~2000ce & 0.57 & $>$ \updated{1.2} & \updated{6.21} & \updated{1.40} & 0.93 & 0.93$-$\updated{1.01} \\[0.11cm]
SN~2012cp & 0.42 & $>$ \updated{1.4} & \updated{2.16} & 0.37 & 0.86 & 0.86$-$\updated{0.91} \\[0.11cm]
SN~2001dl & \updated{0.32} & $>$ 1.8 & \updated{3.73} & \updated{2.08} & 0.91 & 0.91$-$0.93 \\[0.11cm]
SN~2005kc & 0.30 & $>$ 2.1 & \updated{2.96} & \updated{0.74} & 1.25 & 1.25$-$1.28 \\[0.11cm]
SN~2012cu & 1.09 & $>$ \updated{3.1} & \updated{3.22} & \updated{0.32} & 0.90 & 0.90$-$0.97 \\[0.11cm]
SN~2006cm & \updated{1.10} & $>$ 3.4 & 2.97 & \updated{0.74} & 0.91 & 0.91$-$0.99 \\[0.11cm]
SN~2006gj & 0.33 & $>$ \updated{4.2} & \updated{2.01} & \updated{0.52} & 1.55 & 1.55$-$1.58 \\[0.11cm]
SN~2007ss & 0.39 & $>$ \updated{6.3} & \updated{1.93} & \updated{0.44} & 1.12 & 1.12$-$1.13 \\[0.11cm]
SN~1995E	 & 0.77 & $>$ \updated{7.7} & \updated{2.06} & \updated{0.25} & 0.98 & 0.98$-$1.00 \\ [0.11cm]
SN~1998dm & \updated{0.35} & $>$ \updated{8.1} & \updated{2.43} & \updated{1.62} & 0.86 & 0.86$-$0.87 \\ [0.11cm]
SN~2010ev & \updated{0.28} & $>$ \updated{8.1} & \updated{7.61} & \updated{4.61} & 1.14 & 1.14 \\[0.11cm]
SN~2007bm & 0.54 & $>$ \updated{8.5} & \updated{2.07} & \updated{0.59} & 1.17 & 1.17$-$1.18 \\[0.11cm]
SN~1997dt & 0.52 & $>$ \updated{8.6} & \updated{4.86} & \updated{0.38} & 1.15 & 1.15$-$1.16 \\ [0.11cm]
SN~2005bc & \updated{0.43} & $>$ 10.1 & \updated{2.87} & \updated{1.58} & 1.37 & 1.37 \\[0.11cm]
SN~1998bu & 0.38 & $>$ \updated{11.7} & \updated{5.16} & \updated{3.36} & 1.02 & 1.02$-$1.03 \\[0.11cm]
SN~1989B  & 0.39 & $>$ \updated{14.2} & \updated{4.74} & \updated{3.45} & 1.09 & 1.09$-$1.10 \\[0.11cm]
SN~1999gd & 0.45 & $>$ \updated{14.5} & 2.92 & \updated{0.68} & 1.27 & 1.27 \\[0.12cm]
SN~1996ai & 1.75 & $>$ \updated{15.2} & \updated{4.75} & \updated{1.76} & 0.99 & 0.99$-$1.02 \\[0.11cm]
SN~1999ee & \updated{0.31} & $>$ \updated{17.8} & \updated{3.94} & \updated{2.44} & 0.82 & 0.82 \\[0.11cm]
SN~2004ab & 1.64 & $>$ \updated{18.5} & \updated{4.15} & \updated{1.36} & 1.17 & 1.17$-$\refereetwo{1.20} \\[0.11cm]
SN~2003cg & 1.27 & $>$ \updated{40.9} & \updated{3.14} & \updated{0.89} & 1.09 & 1.09 \\
\hline
\end{tabular}                    
\end{table*}

\subsection{Dust distance lower limits}
\label{sec:lower}

For the 33 SNe~Ia showing a constant colour excess, we estimate lower limits on the dust distance. Unfortunately, photometry for many of these SNe is not good enough (e.g. sparse sampling, lack of early-/late-time epochs) to unambiguously exclude dust located at distances inferred in Section~\ref{sec:estimates}. For instance, about half of these SNe have rather poor photometry that results into relatively weak constraints ($\zeta\gtrsim0.3-1$~pc), while SNe with better photometry (e.g. SN~1999e in Fig.~\ref{fig:procedure}) are more constraining. 
\refereetwo{In particular,} SNe with photometry extending to relatively late epochs ($\gtrsim40$~d post peak) are typically characterized by strong constraints on the dust location ($\zeta\gtrsim5$~pc). Again, this highlights the importance of late-time photometry. Nevertheless, comparisons with models suggest that dust for these SNe is likely to be located at pc-scale distances.


\section{Discussion}
\label{sec:discussion}

In Section~\ref{sec:results}, we extracted dust distance values for the 48 SNe~Ia in our sample. The \ebv~evolution for most of the SNe is consistent with dust located at $\zeta\gtrsim1$~pc. In one case, instead, dust is estimated to be much closer to the SN, at distances where CS material connected to the progenitor might be present ($\zeta\sim0.013$~pc, \referee{SN~2003hx}). Here we discuss the implications of our results for explaining the origin of extinction in SNe~Ia. In Section~\ref{sec:comparisons}, we first compare the dust distance values obtained in this study to those available in the literature for some of the SNe in our sample. In Section~\ref{sec:intertellar}, we then provide evidences that dust seen in all 48 SNe is located within IS clouds. In Section~\ref{sec:cloudcloud}, we finally discuss how the ``cloud-cloud collisions'' scenario proposed by \citet{hoang2017} is able to explain our findings within the IS framework and to answer many open questions connected to SNe~Ia suffering extinction by dust.

\subsection{Comparison to previous studies}
\label{sec:comparisons}

\subsubsection{Light echoes}
\label{sec:lightechoes}

An approach to estimate the location of dust in SNe is through the detection of light echoes, namely light scattered by dust into the DLOS \citep[see e.g.][]{patat2005,tylenda2004}. Instead of using early-time photometry as done in this work, light echo measurements rely on imaging observations and focus on later times ($\gtrsim$~1 year). At these epochs, the SN flux along the DLOS has faded considerably and the delayed signal from light scattered by dust produces a ring/arc on the plane of the sky. Simple geometrical arguments are then used to work out the dust location from the time-delay and position of the appeared light echo. This approach was applied to a handful of SNe~Ia in the past, including four objects in our sample: SN~1998bu, SN 1995E, SN~2006X and SN~2014J. 

Two distinct light echoes were detected towards SN~1998bu \citep{cappellaro2001,garnavich2001,drozdov2016}, one estimated to be at $10\pm3$~pc from the SN and the other one between $\sim$~100 and 200~pc. Our lower limit on the dust distance to SN~1998bu is \updated{$\zeta>11.7$}~pc, fully consistent with the outer echo while only marginally with the inner one. Dust located at $207\pm35$~pc was instead inferred from light echo measurements of SN~1995E \citep{quinn2006}, in agreement with our lower limit of \updated{$\zeta>7.7$}~pc. 

A light echo towards SN~2006X was detected by \citet{wang2008b} at $\sim$~300~d after peak, and the dust responsible for the echo \referee{was located at} $\sim27-120$~pc away from the SN. Using images at later epochs ($680$~d), \citet{crotts2008} constrained the same echo to be at $26.3\pm3.2$~pc. Analysis by \citet{drozdov2016} on even later epochs ($\sim1400$~d) resulted in a larger estimate for the SN-dust distance, $80\pm5$~pc. Our calculations suggest that dust responsible for the bulk of extinction seen in SN~2006X is located at \updated{$\zeta=17.3_{-1.0}^{+1.3}$}~pc from the SN, \referee{2.6$\sigma$ away from} the findings of \citet{crotts2008}. Using \lmc~instead of \mw type dust in our models gives \updated{$\zeta=20.1_{-1.4}^{+1.3}$}~pc, in better agreement with the light echo estimate (1.8$\sigma$). 

Light echo detections towards SN~2014J were reported by \citet{crotts2015} and \citet{yang2017}. Specifically, both studies identify an arc-like echo created by dust at $\sim200-300$~pc and a more diffuse echo originated by an extended dust layer with an inner radius of $\sim$~100~pc. Localization of possible dust at distances closer than $\sim$~70~pc were not possible due to difficulties in properly removing the DLOS SN flux within 0.3 arcsec from the SN position. This range includes the location of dust estimated for SN~2014J in this work, \updated{$\zeta=44.6_{-2.0}^{+3.3}$}, highlighting the advantage of our approach to identify dust at relatively nearby distances.

Dust distance estimates presented in this work for SN~1998bu, SN 1995E, SN~2006X and SN~2014J are \referee{compatible to} those inferred for the same SNe from light echo detections. Both approaches exploit the information from light scattered by dust, but light echo studies focus on late-time imaging observations while our technique on photometry at earlier epochs. As shown above, this difference results in our technique being more sensitive to relatively-nearby dust \referee{(within a few tens of pc)} while light echo measurements to relatively-far dust \referee{(distances larger than a few tens of pc)}.




\subsubsection{Time-evolution of polarization}
\label{sec:polarization}

Dust sufficiently close to the SN is predicted to induce time-variability in the polarization signal \citep{patat2005}. The lack of time-variability in the continuum polarization of SN~1986G, SN~2006X and SN~2014J was taken by \citet{patat2015} as evidence that the bulk of extinction suffered by these SNe stems from IS clouds. In the case of SN~2014J, a similar conclusion was reached by \citet{kawabata2014} analysing optical and near-infrared imaging polarimetry. However, observations by \citet{patat2015} and \citet{kawabata2014} were taken at relatively early epochs, between $\sim$~10~d before and $\sim$~35~d after peak brightness. A recent study by \citet{yang2018} analyses imaging polarimetry of SN~2014J at later epochs and claims a variation in the polarization signal occurring 277~d after maximum light. This is interpreted as due to $\sim10^{-6}M_\odot$ of CS dust located at $\sim$~0.2~pc from the SN. Owing to the small mass, this dust shell is expected to contribute only marginally to the extinction \firstit{\citep[see equation 3 in][]{johansson2017}} and therefore not to be captured by the approach presented here.

\subsubsection{Time-evolution of sodium, potassium and calcium}
\label{sec:sodium}

\begin{figure*}
\begin{center}
\includegraphics[width=1\textwidth,trim={0 0 18 0},clip]{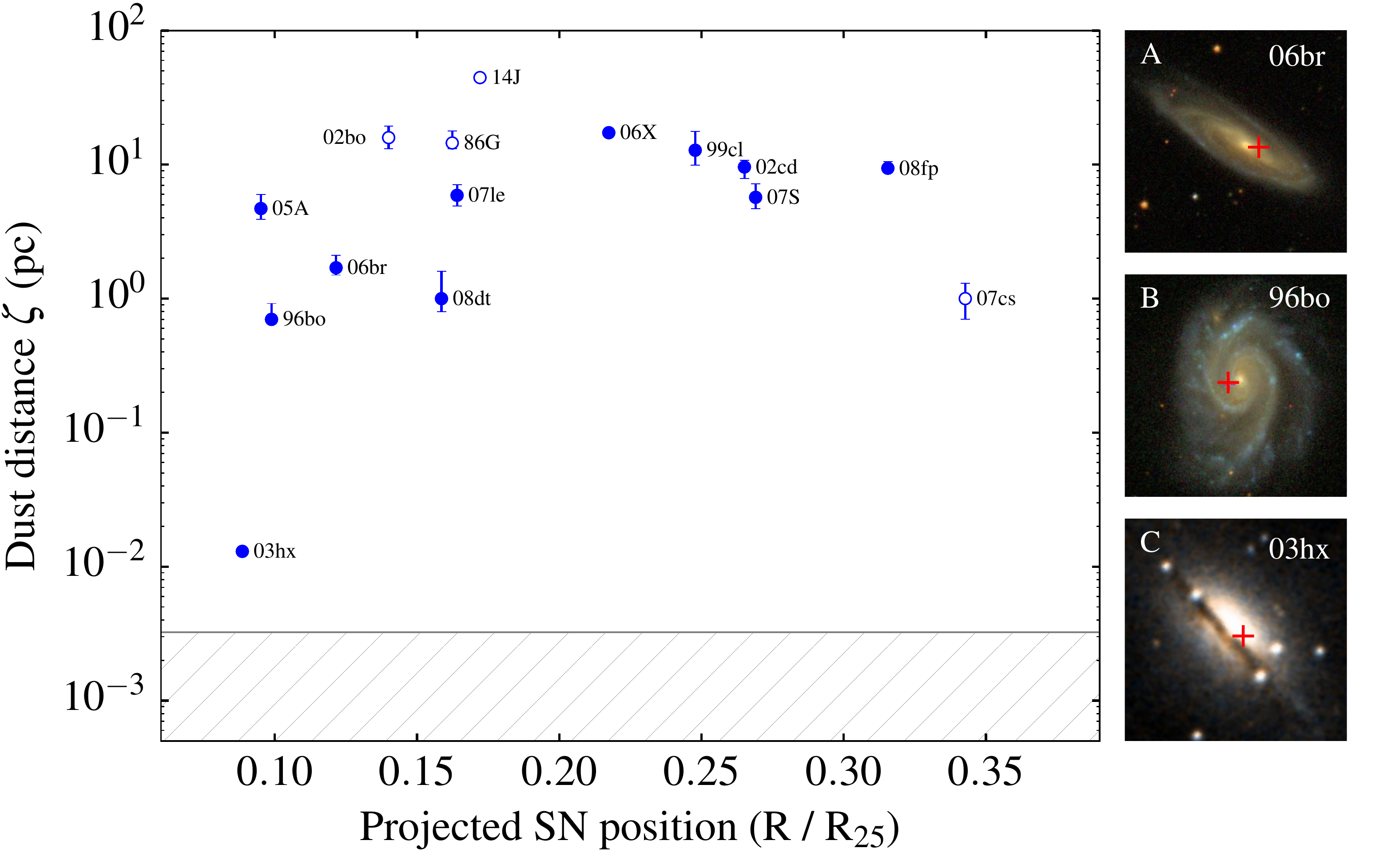}
\caption{Dust distance estimates vs SN position within the host galaxy for the SNe~Ia in our sample showing time-variable \ebv. $R$ is the projected galactocentric distance while $R_{25}$ parametrises the size of the galaxy and is defined as the radius at which the surface brightness of a spiral galaxy falls off to a level of 25 mag/arcsec$^2$ in the B-band. Open symbols denote SNe~Ia occurring within dust lanes likely characterized by higher dust densities. The hashed area corresponds to distances from the SN where dust is unlikely to survive ($\sim$~10$^{16}$~cm, \citealt{amanullah2011}). Right panels show the SN location (red cross) for three events in our sample with relatively small dust distance values: SN~2006br ($\sim$~1.7~pc, panel A), SN~1996bo ($\sim$~0.7~pc, panel B) and SN~2003hx ($\sim$~0.01~pc, panel C).}
\label{fig:galaxy}
\end{center}
\end{figure*}

The detection of time-variable Na~I, K~I and Ca~II features in SNe~Ia can be interpreted as arising in either (i) CS gas subject to ionisation and recombination effects \citep{patat2007} or in (ii) IS gas with covering factor changing with time due to the expansion of the underlying SN photosphere (usually referred to as geometric effects, \referee{\citealt{patat2010}}). 

Time-variable features were found for three SNe~Ia in our sample. Variations in the Na I D profiles were reported for SN~2006X \citep{patat2007} and SN~2007le \citep{simon2009}, while changes in a K I spectral feature were seen in SN~2014J \citep{graham2015,maeda2016}. If attributed to outflowing CS gas (see below), the corresponding distances for the absorbing material would be $\sim$~0.003$-$0.03, 0.1 and 2$-$5~pc for SN~2006X, SN~2007le and SN~2014J, respectively. Our technique places dust at larger distances from the explosion in the case of SN~2006X (17.3~pc), SN~2007le (\updated{5.9}~pc) and SN~2014J (44.6~pc). 

We note that -- with the exception of SN~2006X -- all the detection of time-variable absorption features in the literature are also compatible with geometric effects. In particular, the observed changes in the equivalent widths of these lines are of the order of those induced by the SN photosphere expanding in a fractal patchy IS medium \citep[$\lesssim50$~m\AA~for Na~I~D,][]{patat2010,huang2017}. If attributed to geometric effects, changes observed in SN~2007le and SN~2014J would be qualitatively consistent with our findings of material at 5.6~and 44.6~pc, respectively. For SN~2014J, our estimate would also agree with the one inferred by \citet{maeda2016} from the non-detection of time-variable Na I D (a few $\sim$~10~pc away from the explosion, see their figure 14). For geometric effects to work in the case of SN~2006X, where equivalent widths of Na I D \referee{were} as high as $\sim$~300~m\AA, one would instead need to invoke a highly non-homogeneous IS medium (see also \citealt{chugai2008} and discussion in Section~\ref{sec:cloudcloud}).

\referee{Time-variability was searched but not detected in low-resolution spectra of 9 SNe~Ia in our sample \citep{blondin2009,blondin2017}: SN~1997dt, SN~1998bu, SN~1998dm, SN~1999cl\footnote{Changes in the Na I D features were originally reported for SN~1999cl by \citet{blondin2009}, but later found to be due to measurements errors \citep{blondin2017}.}, SN~2002bo, SN~2003cg, SN~2007S, SN~2007bm and SN~2007ca. Our study suggests that dust is located at interstellar scales ($\gtrsim$~1~pc) for these SNe, and therefore the lack of time-variability can be interpreted as due to the SN photosphere expanding into a rather homogenous interstellar medium (i.e. no detectable geometric effects).}


\subsection{Evidence for interstellar extinction}
\label{sec:intertellar}

In Section~\ref{sec:results}, we analysed the \ebv~evolution for all the SNe~Ia in our sample and using dust models \referee{inferred} the location of dust in each of them. Here we discuss what these numbers tell us in terms of the origin of the extinction.
\firstit{In particular, we frame our discussion in terms of whether the dust responsible for the extinction is connected (CS) or not connected (IS) with the SN progenitor system}.


\ebv~curves consistent with no-time variability are found for 33 SNe in our sample and the dust inferred to be at distances larger than $\sim$~0.3~pc (Table~\ref{tab:results2}). As discussed in Section~\ref{sec:lower}, however, photometry for about half of these SNe is rather poor and does not extend to late-enough epochs to place strong constraints on the dust location. SNe with good photometry, e.g. those with last epoch taken later than 40~d after maximum, are more constraining and associated with dust at $\zeta\gtrsim5$~pc .
These distances are consistent with the extinction arising within the IS medium from dust not connected to the SN progenitor system.

\ebv~curves showing time-variability are found for 15 SNe in our sample (Table~\ref{tab:results1}). For most of these SNe, our simulations indicate that dust is located between 1 and 50~pc and therefore that the origin of the observed extinction is likely within IS clouds. SN~2003hx is instead predicted to have dust much closer to the explosion (0.013~pc). Given the proximity of dust, it is tempting to think that the extinction for this SN stems from CS material \firstit{associated to the progenitor system}. However, the proximity of dust might also be reflecting extinction by IS clouds in a relatively high-density region of the host galaxy. In the following, we try to investigate this in more details by looking at the location of this SN.


The location of SN~2003hx is interesting in the sense that this SN occurred very close to the center of the host galaxy. This is shown in Fig.~\ref{fig:galaxy}, where we plot the dust distance estimates vs the SN location for SN~2003hx and the remaining 14 SNe with time-variable \ebv.
The SN position is calculated as $R/R_{25}$, where $R$ is the projected galactocentric distance and $R_{25}$ parametrises the size of the galaxy and is defined as the radius at which the surface brightness of a spiral galaxy falls off to a level of 25 mag/arcsec$^2$ in the B-band. Values for $R_{25}$ are taken from the Asiago Supernova catalogue \citep{barbon1999}. Although $R/R_{25}$ probes a projected rather than a real distance to the galactic centre, Fig.~\ref{fig:galaxy} shows that SN~2003hx not only is the object in our sample with dust closer to the explosion but also the one that exploded closer to the center of the galaxy ($R/R_{25}\lesssim0.1$). 

A potential trend of increasing dust distance with increasing galactocentric radius is found when looking at the other 14 SNe. This is especially true when removing SN~2002bo, SN~2007cs, SN~1986G and SN~2014J that are known to occur in peculiar regions of their host galaxy, specifically within dust lanes. In particular, SNe at intermediate distance from the galactic centre ($0.1\lesssim R/R_{25}\lesssim0.2$) are associate with dust between $\sim$~1 and $\sim$~5~pc, while those at larger distances ($R/R_{25}\gtrsim0.2$) with dust around 10~pc. The observed trend is what one would expect if the extinction in all the 15 SNe is located within IS clouds, with the chances of interacting with a dust layer increasingly closer to the SN site becoming higher as one moves from the outskirts to the center of the galaxy.

We conclude that the extinction in \textit{all} the 48 SNe~Ia is likely to originate from dust in the IS medium not connected to the SN progenitor system.

\begin{figure*}
\begin{center}
\includegraphics[width=0.78\textwidth]{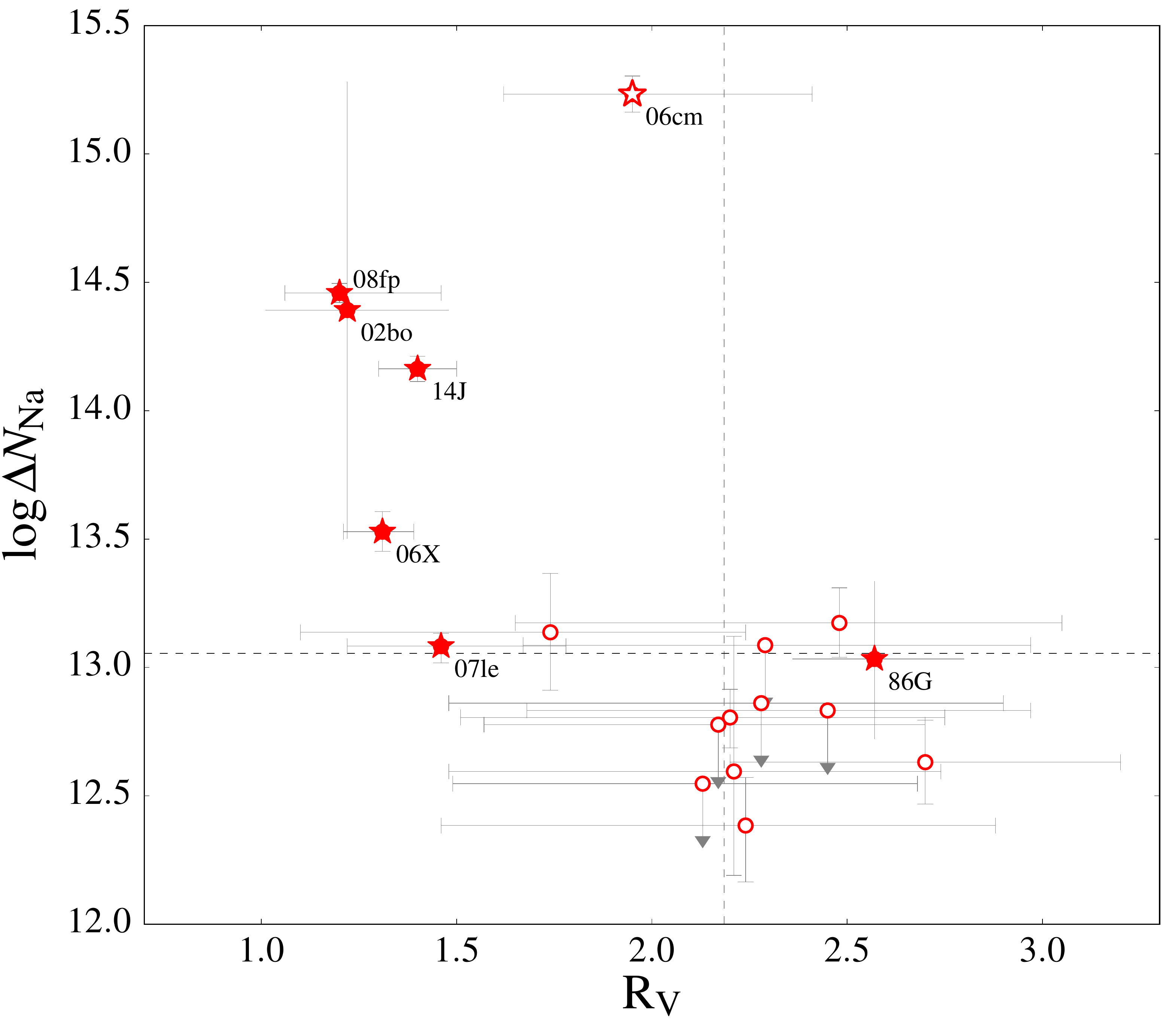}
\caption{Sodium column-density excess, $\Delta N_\mathrm{Na}$, vs \rv~values for 18 SNe in the literature \citep{phillips2013,amanullah2015,ritchey2015}. Sodium excess are calculated as in equation~(\ref{eq:sodium}) following \citet{phillips2013}. SNe in our sample are labelled and shown with star symbols\referee{: filled stars are SNe with time-variable \ebv~and a dust distance estimate, while open star corresponds to SN~2006cm that shows a constant \ebv~evolution and for which a lower limit on the dust distance is placed. SNe that are not in our sample are shown with open circles}. The horizontal (vertical) dashed line divides the sample so that half of the points are above (to the right of) and half below (to the left of) this line.}
\label{fig:sodium}
\end{center}
\end{figure*}

\subsection{A scenario for interstellar dust}
\label{sec:cloudcloud}

The extinction observed for the 48 SNe~Ia in our sample is likely to occur within IS clouds. Here we discuss a possible scenario in the IS framework that can account for most of the open questions connected to reddened SNe~Ia, namely the origin for the low \rv~values, the unusually large sodium abundance, the preference for blue-shifted absorption features and the time-variability of sodium and potassium features (see Section~\ref{sec:introduction}). These open questions, together with their possible interpretation within the IS scenario, are summarised in Table~\ref{tab:interpretation} and discussed in the following.

In a recent study, \citet{hoang2017} suggested that strong radiation pressure from a SN explosion can accelerate dust grains to high terminal velocities and induce collisions between the cloud closest to the SN and another cloud in its vicinity. Specifically, grains of size $a$ at distance $\zeta$ from the SN will be accelerated to velocities
\begin{equation}
v \sim 171\,\bigg(\frac{L_\mathrm{bol}}{10^8 L_\mathrm{\odot}}\bigg)^{1/2}\bigg(\frac{\zeta}{100~\mathrm{pc}}\bigg)^{-1/2}\bigg(\frac{a}{10^{-5}~\mathrm{cm}}\bigg)^{-1/2}\mathrm{km~s^{-1}}
\label{eq:vel}
\end{equation}
where $L_\mathrm{bol}$ is the bolometric luminosity of the SN (see also \citealt{hoang2015}). Hence, even relatively distant dust clouds ($\zeta\sim100$~pc) can be accelerated to rather large velocities ($\sim200$~km~s$^{-1}$) for bolometric luminosities of the order of $L_\mathrm{bol}\sim10^8\,L\mathrm{\odot}$. For a typical SN~Ia, this luminosity is reached within a few days after explosion ($\sim$~15~d before peak, \citealt{pereira2013}).

The peculiarly low \rv~values seen in some SNe~Ia are explained by this scenario in terms of a shift in grain sizes from larger to smaller values caused by the cloud-cloud collisions \citep{hoang2017}. In other terms, it is the SN itself that affects the IS environment and causes the dust properties to change from \referee{known} (i.e. MW$-$like) to peculiar. Supporting this picture are observations by \citet{hutton2015} which indicate that dust at the location of SN~2014J in M82 was normal  ($R_\mathrm{V}\sim3$, see their figure 6) a few years before explosion.

Moreover, the cloud-cloud collision scenario holds promises for explaining the Na column-density excess observed in some reddened SNe~Ia. Following \citet{phillips2013}, this can be defined as
\begin{equation}
\Delta N_\mathrm{Na} = N_\mathrm{Na} - 10^{13.180 + 1.125 \log A_V}~~,
\label{eq:sodium}
\end{equation}
where $N_\mathrm{Na}$ is the observed sodium column density while the second term the one expected from the extinction $A_V$ if the dust-to-gas relation seen in the MW is valid. In the scenario proposed by \citet{hoang2017}, the collisions between clouds not only leads to a change in the grain size distribution but also enhances the abundance in the gaseous phase of sodium (and other atoms), thus explaining why Na column densities are unusually strong in some reddened SNe~Ia.

\begin{table*}
\centering
\caption{Observational constraints for reddened SNe~Ia, together with their possible interpretations within the IS scenario.}
\label{tab:interpretation}
\begin{tabular}{ll}
\hline
Observation & IS scenario interpretation  \\
\hline
Low \rv~ & SN radiation $\rightarrow$ cloud-cloud collisions $\rightarrow$ from big to small dust grains \citep{hoang2017}  \\
Unusually strong sodium & SN radiation $\rightarrow$ cloud-cloud collisions $\rightarrow$ enhancement of atoms (e.g. sodium) in gaseous phase \\
Blue-shifted absorption features & SN radiation $\rightarrow$ clouds accelerated towards the observer (see equation \ref{eq:vel}) \\
Variable absorption features & Expanding SN photosphere / geometric effect \citep{chugai2008,patat2010}\\
\hline                           
\end{tabular}                                   
\end{table*}

Given the above arguments, a correlation might be expected between \rv~and $\Delta N_\mathrm{Na}$ in reddened SNe~Ia. In Fig.~\ref{fig:sodium}, we show these two parameters for a sample of 18 SNe~Ia\footnote{Note that here we include only SNe~Ia from the literature displaying a Na excess, i.e. those with $\Delta N_\mathrm{Na}>0$.} from the literature \citep{phillips2013,amanullah2015,ritchey2015}. 
While we use values taken from the literature for most of the SNe, $A_V$ for the 7 SNe that are also in our sample are calculated using the \ebvdlos~from Tables~\ref{tab:results1} and \ref{tab:results2}. We do not see a clear correlation between \rv~and $\Delta N_\mathrm{Na}$ in Fig.~\ref{fig:sodium}. However, we notice a preference for SNe~Ia with lower \rv~to have larger $\Delta N_\mathrm{Na}$. It is intriguing that most of the SNe characterized by low \rv~and high Na abundances are those in our sample for which we estimate the dust to be at pc-scale distances to the explosion (SN~2002bo, SN~2006X, SN~2007le, SN~2008fp and SN~2014J). This should be a natural prediction in the scenario proposed by \citet{hoang2017}, where the efficiency of the collision mechanism is expected to be higher for closer compared to more distant dust clouds. 

The acceleration of dust clouds towards the observer can also explain the observed preponderance of blue-shifted absorption features in the spectra of some reddened SNe~Ia. In particular, velocities estimated from equation~(\ref{eq:vel}) show a reasonable match to those observed in SNe~Ia ($\sim$~100~km~s$^{-1}$). It is also encouraging that blue-shifted absorption features were identified in almost all the events in Fig.~\ref{fig:sodium} showing low \rv~and high sodium excess, namely SN~2002bo, SN~2006X, SN~2006cm, SN~2007le, SN~2008fp and SN~2014J \citep{sternberg2011,foley2012,phillips2013,graham2015,maeda2016}. \citet{phillips2013} already pointed out that the SNe in their sample with unusually strong Na are all associated with blue-shifted absorptions. In contrast to the interpretation provided here, however, this was taken as evidence for CS material \referee{since the presence of outflowing material connected to the progenitor system was the only natural explanation for the observed blue-shifted features\footnote{Note that the scenario of \citet{hoang2017} had not been proposed yet at the time the study by \citet{phillips2013} was conducted.}.}

As discussed in Section~\ref{sec:comparisons}, the observed time-variabilities in Na and K features can only be explained in the IS framework by invoking some geometric effects. While most of the variations seen in SNe~Ia are compatible with the projected SN photosphere expanding in a fractal patchy IS medium \citep{patat2010,huang2017}, those observed in SN~2006X would require a highly non-homogeneous IS medium \citep[see also][]{chugai2008}. It is again encouraging that all the normal SNe~Ia showing time-variability in either Na or K (SN~2006X, SN~2007le, SN~2014J) have dust in relatively nearby interstellar clouds according to our analysis, at distances where the covering factor of IS clouds can be significant and rapidly varying with time.

In summary, our analysis suggests that low \rv~values, unusually strong Na abundances and blue-shifted and time-variable absorption features are likely to originate from IS clouds relatively nearby to the explosion (pc-scale distances) and experiencing collisions induced by the SN radiation pressure \citep{hoang2017}.

\section{Conclusions}
\label{sec:conclusions}

We have studied the temporal evolution of the colour excess \ebv~in 48 SNe~Ia. Comparing the observed evolution to that predicted by dust models from \citet{bulla2018}, we constrained the location of dust $\zeta$ in each SN. We summarise the main findings of our analysis in the following:

\begin{itemize}[leftmargin=*]

\item 15 SNe in our sample show time-variable \ebv. We predict dust to be located at distances in the range \mbox{$0.013-1$}~pc for 4 of these SNe, while between $\sim1$ and 50~pc for the remaining SNe;

\item 33 SNe in our sample are consistent with no time-variability in \ebv. For these SNe, we place lower limits on the dust distance that range from $\zeta\gtrsim$~0.3 to $\zeta\gtrsim$~40~pc. We highlight how the rather poor light-curve sampling, and specifically the lack of late-time epochs, in about half of these SNe result in poor constraints on the dust location;

\item The inferred dust locations for all the 48 SNe~Ia are consistent with the observed extinction arising within IS clouds. This is corroborated by the fact that  SNe interacting with dust relatively nearby ($\zeta\lesssim1$~pc, at possible CS scales) are those in our sample that are the closest (in projection) to the center of their host galaxy, where high-density dust environments are expected;

\item Comparing our results to those in the literature, we find that SNe interacting with interstellar dust at pc-scale distances show a preference for low \rv, unusually strong Na absorption and blue-shifted and time-variable absorption features;

\item In the IS framework, we interpret the last point as a possible evidence supporting the cloud-cloud collision scenario by \citet{hoang2017}. Relatively nearby IS clouds can be accelerated towards the observer (hence blue-shifted absorption) by strong SN radiation pressure and then collide with clouds in their vicinity, thus causing the destruction of big into smaller grains (hence low Rv) and the enhancement of sodium in the gaseous phase (hence unusually Na absorption). \referee{In the IS scenario, the observed time-variabilities of sodium and potassium would instead be explained by the change in the covering factor of the IS gas following the expansion of the underlying SN photosphere (namely by invoking geometric effects);}

\item We do not find any evidence for CS dust, thus tentatively supporting progenitor scenario with rather clean local environments (e.g. double-degenerate systems).

\end{itemize}

In this study, we highlighted the power of photometry to place constraints on the location of dust towards SNe~Ia, showing in particular the importance of both early- and late-time data. Surveys like the ongoing Zwicky Transient Facility (ZTF, \citealt{kulkarni2016}) will be valuable in finding reddened and nearby SNe~Ia at early epochs, allowing for dedicated photometric follow-ups to sample the light-curves of these SNe in the months following the explosion. This will allow us to test the results presented here on an even larger sample, possibly solving the questions about the origin of the extinction in SNe~Ia.

\section*{Acknowledgements}

\refereetwo{The authors are thankful to the anonymous reviewer for his/her valuable comments}. This research has made use of the CfA Supernova Archive, which is funded in part by the National Science Foundation through grant AST 0907903. The authors acknowledge support from the Swedish Research Council (Vetenskapsr\aa det) and the Swedish National Space Board.


\bibliographystyle{mn2e}
\bibliography{bulla2018b}

\end{document}